\documentclass[12pt]{iopart}

\newcommand{\be}{\begin{eqnarray}}
\newcommand{\ee}{\end{eqnarray}}

\newcommand{\nbar}{\bar{n}}

\newcommand{\caf}{\ensuremath{^{40}{\rm Ca}^{+}\, }}

\usepackage{latexsym}
\usepackage{graphics}
\usepackage[squaren,thickspace,thickqspace]{SIunits}
\usepackage{multirow}
\usepackage{textcomp}

\begin{document}

\title{Cooling atomic ions with visible and infra-red light}

\author{F.~Lindenfelser, M.~Marinelli, V.~Negnevitsky, S.~Ragg and J.~P.~Home}

\address{Institute for Quantum Electronics, ETH Z\"urich, Otto-Stern-Weg 1, 8093 Z\"urich, Switzerland}

\ead{jhome@phys.ethz.ch}

\begin{abstract}
We demonstrate the ability to load, cool and detect singly-charged calcium ions in a surface electrode trap using only visible and infrared lasers for the trapped-ion control. As opposed to the standard methods of cooling using dipole-allowed transitions, we combine power broadening of a quadrupole transition at 729\,nm with quenching of the upper level using a dipole allowed transition at 854\,nm. By observing the resulting 393\,nm fluorescence we are able to perform background-free detection of the ion. We show that this system can be used to smoothly transition between the Doppler cooling and sideband cooling regimes, and verify theoretical predictions throughout this range. We achieve scattering rates which reliably allow recooling after collision events and allow ions to be loaded from a thermal atomic beam. This work is compatible with recent advances in optical waveguides, and thus opens a path in current technologies for large-scale quantum information processing. In situations where dielectric materials are placed close to trapped ions, it carries the additional advantage of using wavelengths which do not lead to significant charging, which should facilitate high rate optical interfaces between remotely held ions.
\end{abstract}

\maketitle

\section{Introduction}
\label{intro}

Trapped ions are among the leading candidates for quantum information processing and quantum simulation, as well as being leading contenders for realizing accurate atomic clocks. In all areas, scaling to a larger number of ions is desirable. For quantum computing, two primary methods are envisioned for scaling. The first involves the shuttling of ions in an multi-zone trap array \cite{98Wineland2, 02Kielpinski}, for which microfabricated traps with large numbers of segmented electrodes are required. The second method involves linking remote ion traps via optical interfaces. This can either be performed using free-space photons or using high-finesse optical cavities. Scaling using both methods would greatly benefit from the use of integrated optical components within ion trap arrays. Integrated waveguides were recently used to address ions in a micro-fabricated trap using laser light at 670\,nm \cite{16Mehta}, but further work is required before all-optical delivery can be incorporated into trapping structures. For this purpose, it seems desirable to work at wavelengths in the visible and Infra-Red (IR) regions of the spectrum, where technologies for integrated optics are relatively mature.

A second reason to work with longer wavelengths rather than the ultraviolet light commonly used in most ion trap experiments is that the charging of non-conducting materials used for optical interfaces or appearing due to  contamination of nearby electrode surfaces is much reduced. This has been studied in the context of trap chips~\cite{10Harlander}, and has repeatedly been a major concern when integrating ions with optical cavities \cite{13Steiner}. Atomic ions are highly susceptible to charges on nearby surfaces, which displace the ion from the centre of the trapping region leading to excess micromotion~\cite{98Berkeland2}. Though this can be tolerated at some level in quantum information processing, it leaves the ions susceptible to fluctuations in the RF power driving the trap. Other applications in quantum control place critical requirements on micromotion - one example is the combination of ion traps with cold neutral atoms, where the high-energy micromotion produces unwanted heating \cite{12Schmid,11Zipkes,12Cetina,12Schneider}.
In anharmonic trapping potentials, the displacement due to stray fields also leads to undesirable shifts in the trap frequency~\cite{11Home}. Avoiding light scattering on nearby electrodes and dielectrics becomes particularly difficult as ion traps are reduced in size, because the ion-electrode distance is reduced.

\begin{figure}
\centering
\includegraphics{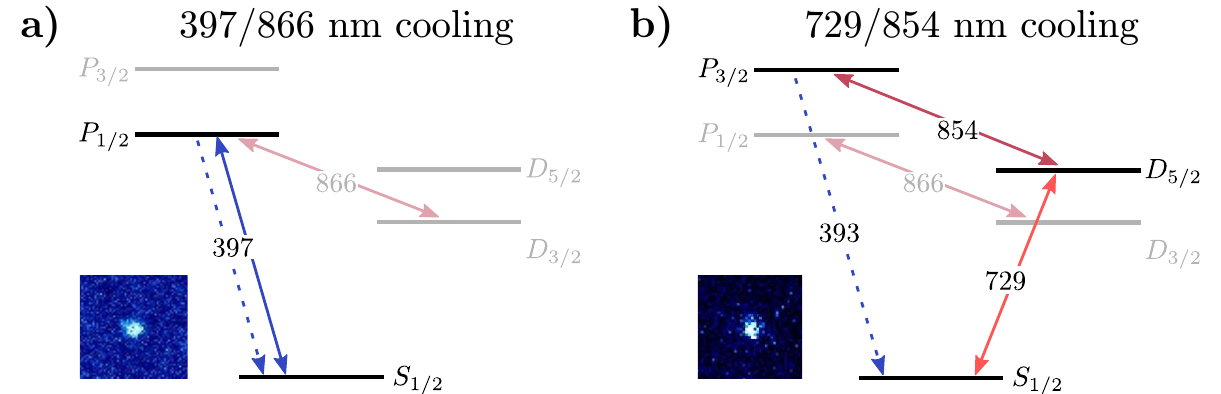}
\caption{Level diagram of \caf to illustrate the traditional scheme for driving ion fluorescence (a) and the one presented here (b). Lifetimes of $P$ levels are on the order of 10\,ns, lifetimes of $D$ levels on the order of 1\,s. Decays from $P_{3/2}$ to $D_{5/2}$ or $P_{1/2}$ to $P_{3/2}$ happen with a branching ratio of 7\,\%. The laser at 866\,nm is needed for repumping population which might get trapped in $D_{3/2}$. Insets show CCD camera images when illuminating \caf with the indicated lasers. Detection time (a): 0.5\,s, (b): 2\,s. Although the fluorescence picture in (b) is taken at longer detection time we still detect close to no background scattered light.}
\label{fig:Caschem}
\end{figure}

The primary laser cooling technique used for trapped atomic ions is Doppler cooling. For commonly-used species, a dipole-allowed transition between the ground $S_{1/2}$ and excited $P_{1/2}$ states with a natural linewidth of between $14.5$\,MHz and $21.6$\,MHz is driven using laser light detuned from the transition resonance. Doppler cooling works in the weak-binding limit \cite{79Wineland}, in which the primary motional sidebands of the transition sit within the lineshape of the transition. In this limit ions can be cooled over a large range of frequency detunings including those which arise from Doppler shifts experienced by ions due to motion in the trap. Most ion trap experiments work with secular frequencies $<10$\,MHz, and thus for dipole-allowed transitions this weak-binding approximates the cooling well (a small number of experiments have worked at higher frequencies in order to perform resolved-sideband cooling with similar transitions \cite{95Jefferts, 07BrownPrivate}). The atomic structure of commonly-used atomic ions means that the dipole transition from the internal ground state is in the ultraviolet region of the spectrum (the exception being barium, for which this transition is in the visible blue at 493\,nm). In barium, strontium, calcium and ytterbium an additional complication is the presence of $D$ levels which lie at lower energy than the $P$ state, requiring the use of repumping lasers to return the ion into the $S-P$ manifold (see Figure\,\ref{fig:Caschem} a)). However these $D$ levels also present a useful feature, namely long-lived levels which have been used successfully as optical qubits \cite{99Roos,16Mehta}. These are connected to the ground state via quadrupole transitions, which have very small natural linewidths - for example in calcium this is 0.14\,Hz.

In this paper we describe the operation of an ion trap with no UV lasers, in particular showing the ability to load and re-cool hot calcium ions with all cooling performed by IR and visible laser beams at 729\,nm, 854\,nm and 866\,nm (see Figure\,\ref{fig:Caschem}\,b)). We combine this with lower intensity settings for the same lasers in order to reach the quantum ground state of motion for a motional mode at 2.7\,MHz by driving the same transitions. In addition we use the ability to tune the intensity to achieve both Doppler and resolved-sideband cooling, to study the crossover between these two regimes and find good agreement between experiment and the theory of Cirac et al. \cite{92Cirac}. The trap is operated under conditions comparable to other current trapped-ion setups for quantum information. Our work provides a significant extension to two recent papers which demonstrated Doppler cooling using IR and visible radiation. The first uses the same transitions as in our work, but the trap frequencies of 560\,kHz and 950\,kHz \cite{08Hendricks} are significantly lower, and no sideband cooling is performed. The second, by Zhou et al. \cite{13Fei} used 732\,nm and 866\,nm radiation to drive fluorescence using the quadrupole transition between the ground state and the low-lying $D_{3/2}$ level, with low axial-mode trap frequencies between 90 and 470\,kHz. 
In both of the prior works, the repumping light at 854\,nm and 866\,nm was off-resonant. Although this simplifies the theoretical description, it also limits the scattering rates and cooling capture range that can be achieved. In this article we first present a study of cooling dynamics under the influence of the 729\,nm, 854\,nm and 866\,nm light with the 854\,nm laser tuned close to the zero-field resonance. We then show that, contrary to the previous work which required sequences in which periods of cooling with the 397\,nm and 866\,nm lasers were interspersed with the quadrupole transition cooling in order to maintain stable trapping, we can use the studied cooling scheme to operate our trap with all laser light above 700\,nm. In particular we show the ability to cool hot ions, such as is required for us to demonstrate loading and to recover from background gas collisions.

\section{Methods and simulations}
\label{sec:exptmethods}

\subsection{Calcium ion trapping}
We trap and cool a single calcium-40 ion in a surface-electrode ion trap with an ion-electrode distance close to $90\,\mu$m. The trap has been described in detail elsewhere~\cite{15Lindenfelser}. All laser beams are directed parallel to the surface of the ion trap, while the magnetic field of $4.05$\,Gauss is directed perpendicular to the surface. The key element allowing us to access the Doppler cooling regime on the quadrupole transition is the high 729\,nm radiation intensity addressing this transition, which is achieved by focussing up to 45\,mW of light to a spot-size of $6.3\,{\mu}$m$\,\times\,9.2\,{\mu}$m, which is close to the diffraction limit imposed by the vacuum windows in our setup (these are 1.5\,cm in diameter at a distance of 8.9\,cm). This allows intensities of up to 494\,W/mm$^2$ at the ion. The 854\,nm laser which quenches this transition (as well as the 866\,nm laser light which repumps from the $D_{3/2}$ state) is directed counter-propagating with the 729\,nm laser. The secular motion of the ion in the surface trap consists of oscillations at 2.3\,MHz, 2.7\,MHz and 5\,MHz, of which the highest frequency mode oscillates at approximately $87$\,degrees to the electrode plane. The other two modes are approximately parallel to the electrode surface, directed at 43 and 47\,degrees to the 729\,nm laser beam.

Quantum state discrimination between the $S_{1/2}$ and $D_{5/2}$ levels can be performed using the standard approach using the cooling lasers at 397\,nm and 866\,nm. The detection probability for a single emitted photon at 397\,nm is $0.0028(2)$, measured by using the ion as a single-photon source, by first shelving and then de-shelving the ion from the $D_{3/2}$ level using the 866\,nm laser \cite{10Shu}. 
State detection is performed in a standard window of 350\,$\mu$s, during which we detect $20$ photons for an ion in the $S_{1/2}$ level and $0.8$ photons for an ion in the $D_{5/2}$ level (the latter is primarily due to background scattering from the electrodes of the trap).
Our apparatus' detection efficiency of 393\,nm fluorescence is measured by first transferring population from $S_{1/2}$ to the $D_{5/2}, M_J = -5/2$ state using a resonant coherent pulse of light from the 729\,nm laser at low power. We then detect to confirm that the ion was excited. If no fluorescence light is detected we assume the population to be in $D_{5/2}$ and subsequently we record whether a photon is detected when repumping the ion using 854\,nm light, with the primary scattering path being via the $P_{3/2}, M_J = -3/2$ level. We thus obtain a photon detection efficiency of 0.0044(1). This is higher than for 397\,nm light since our imaging system is aligned with the magnetic field and is mostly sensitive to circularly polarized light which is predominantly emitted when de-shelving $D_{5/2}, M_J=-5/2$. 

\subsection{Measurements}
Although the primary aim of the experimental work was to demonstrate the ability to operate the trap with IR and visible light, the experiment also offers the possibility to probe laser cooling with the same set of laser systems in both the weak-binding and strong-binding limits. This allows us to check our understanding of the underlying physics, and therefore to make predictions regarding extensions to our work. To do this, we compared theory to experiment over a range of settings of the powers of the 729\,nm and 854\,nm lasers. The experiments allowed us to extract the cooling rate and final temperature, and to compare that to the spectrum of 393\,nm fluorescence.

In order to measure the motional state of the ion, we first repump the ion to $S_{1/2}, M_J = -1/2$, and subsequently apply a weak probe pulse of 729\,nm light resonant with the blue motional sideband of the transition to the $D_{5/2}, M_J = -5/2$ state. Making use of frequency selectivity, we spectoscopically pick out the sideband of the secular mode of oscillation at 2.7\,MHz. The probability to find the ion in the $S_{1/2}$ level as a function of the duration $t_p$ of the probe pulse follows the functional form
\be
P(S) = a + (1-a) \sum_{n = 0}^{\infty} p(n) \frac{1}{2}\left(1 + e^{-\gamma t_p} \cos(\Omega_{n} t_p \right) 
\ee
which depends on the motional probability $p(n)$ to be in the $n$th motional energy eigenstate prior to the probe pulse, and the Rabi frequency for the sideband transition given this starting state $\Omega_n = \sqrt{n+1} \Omega_0$. The decay rate $\gamma$ is a phenomenological parameter which we use to account for decoherence, which we think is primarily due to intensity fluctutations and $a$ accounts for state preparation and measurement errors. In order to extract the mean thermal occupancy $\bar{n}$, we assume a thermal distribution  $p_{\bar{n}}(n) = \bar{n}^{n}/(\bar{n} + 1)^{n + 1}$. 
In fitting data, $\bar{n}$, $\Omega_0$, $\gamma$, and $a$ are floated.
\\
\\
Measurements of fluorescence from the 729/854\,nm driven ion are performed after cooling it to the quantum ground state. This cooling is performed using continuous sideband cooling with low 729\,nm and 854\,nm light intensities \cite{99Roos}. The settings for the 729\,nm and 854\,nm lasers are then changed to those at which we want to probe the scattering rate, and 393\,nm fluorescence is collected for a fixed duration of $200\,\mu$s. This method (in particular the short duration) is chosen because the spectrum changes as the ion heats up, due to the increased coupling on the motional sidebands. For longer probe pulses this complicates the analysis and makes it challenging to reliably extract parameters such as laser intensities and detunings from the data.

\subsection{Simulations of cooling}
In order to gain a better understanding of the cooling throughout the different regimes of laser powers, we use the theory of Cirac and co-workers \cite{92Cirac}. This theory is valid within the Lamb-Dicke regime, and uses the separation in timescales between the relatively fast return of the internal state levels to the steady state versus the slower timescales on which the motional states evolve. This allows cooling and heating rate coefficients $A_+$ and $A_-$  to be calculated from a sum of contributions from the fluctuation spectrum of the two-time correlation function of the dipole force and from a diffusion term, where both are evaluated using the steady-state density matrix for the internal electronic state of the atom. From the rate coefficients, the cooling rate $\Gamma_c = A_- - A_+$ and the steady state motional occupancy $\bar{n}_\infty = A_+/(A_--A_+)$ can be calculated.
The theory is valid in both the weak-binding and strong-binding limit, and allows a description of the cooling beyond simple lineshapes such as those produced by two and three-level approximations. The Lamb-Dicke assumption is a good approximation for our experiments, where the Lamb-Dicke parameters for the two transitions are  $\eta_{\rm 729} = 0.039$ and $\eta_{854} = 0.033$, and values of $\bar{n}$ which we consider to be around $<10$ quanta.

\begin{figure}
	\centering
	\includegraphics{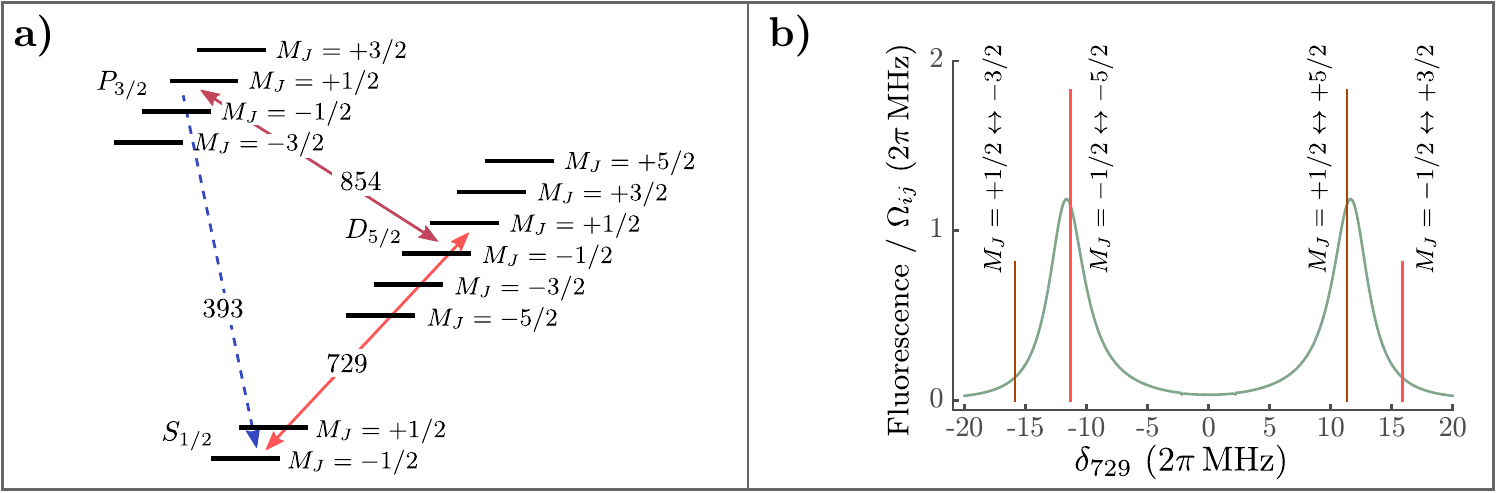}
	\caption{a) Zeeman sublevels of \caf relevant for the 729/854\,nm driving scheme. b) Possible transitions between Zeeman sublevels of $S_{1/2}$ and $D_{5/2}$ driven by the 729\,nm laser at our polarization and direction settings versus the detuning of their resonance frequency at $B=4.05$\,Gauss. The height of vertical lines indicates their driving strength $\Omega_{ij}$ between the Zeeman levels for $\Omega_{729} = 2\pi\times2$\,MHz. The solid line is a simulation of fluorescence at $\Omega_{729} =2\pi\times2$\,MHz, $\Omega_{854} = 2\pi\times4$\,MHz and our experimental settings.  }
	\label{fig:transitions}
\end{figure}

To calculate the cooling dynamics, we first solve for the steady state of the internal electronic states under the influence of the laser fields. We simplify this numerically by neglecting the influence of the $D_{3/2}$ level, and thus solve the Optical Bloch Equations (OBEs) of the 12 states of the $S_{1/2}$, $P_{3/2}$ and $D_{5/2}$ levels. Since the 729\,nm light which we use to excite the transitions is single-frequency and linearly polarized at $90$\,degrees to the magnetic field, the transitions which can be driven are the four $\Delta M_J = \pm 2$ transitions, for which the frequencies relative to the zero-field frequency of the $S_{1/2} \leftrightarrow D_{5/2}$ transition are shown in figure\,\ref{fig:transitions}. In the cooling experiments, we primarily work with zero detuning of the 854\,nm laser and the 729\,nm laser tuned close to the lower-frequency transitions. For the powers used, an ion starting in the $S_{1/2}, \ M_J = -1/2$ state will primarily cycle between this state and the $D_{5/2}, \ M_J = -5/2$ and $P_{3/2}, M_J = -3/2$ states. The ion has a 3\,\% probability to exit from this closed loop through decay from $P_{3/2}, M_J = -3/2$ to $D_{5/2}, \ M_J = -3/2$ or $D_{5/2}, \ M_J = -1/2$, or via the $D_{3/2}$ state. After repumping from these states, ions have a significant probability of ending up in the $S_{1/2}, M_J = +1/2$ state. From simulations of the OBE for typical parameters of the 854\,nm and 866\,nm lasers we find that the probability to exit the cooling cycle for an ion starting in the $D_{5/2}, M_j = -5/2$ level is between $0.6\%$ and $1.5\%$, dependent on the 854\,nm laser intensity (high intensity gives a higher probability of exiting the cooling cycle). Repumping back into the cooling cycle is dominated by the $S_{1/2}, M_J = +1/2 \rightarrow D_{5/2}, M_J = -3/2$ transition which is close to resonance with the 729\,nm laser light, and then takes place via the $P_{3/2}, M_J = -1/2$ state.  The resonant conditions which are involved in both the cooling cycle and the repumping lead to an interplay between the two which strongly influences the steady-state lineshapes of the atom as a function of the laser detuning, such that the fluorescence is typically peaked between these two transitions rather than at one or the other resonance.

\section{Experimental results}

\subsection{Fluorescence and cooling}
Measurements of the fluorescence were performed for a variety of settings of the intensity and detuning of the 729\,nm and 854\,nm lasers, controlled by the drive powers of Acousto-Optic-Modulators (AOMs) in the optical beam paths. Care was taken to account for changing diffraction efficiencies of the AOMs over the frequency scan range. Fits to fluorescence curves as a function of the 729\,nm laser detuning $\delta_{\rm 729}$ allowed the extraction of the Rabi frequencies for both lasers $\Omega_{\rm 729}$, $\Omega_{\rm 854}$ and the detuning of the 854\,nm laser $\delta_{\rm 854}$, which were floated parameters for the fit. Additional parameters which were included in the model were the magnetic field strength and the frequency of the $S_{1/2}\leftrightarrow D_{5/2}$ transition at zero magnetic field. The modelled linewidth of the 854\,nm laser was fixed at $\gamma_{\rm 854} = 2 \pi \times 500$\,kHz while that of the 729\,nm laser was fixed at $\gamma_{\rm 729} = 2 \pi \times 10$\,Hz. Experimentally, the 854\,nm laser linewidth is specified to be $<500$\,kHz, and the 729\,nm could vary up to 2\,kHz linewidth depending on the settings of various locking parameters. Although these linewidths are not well known from day to day in our experiments, we find in the simulations that neither plays a significant role in the results presented below. 

\begin{figure}
	\centering
	\includegraphics{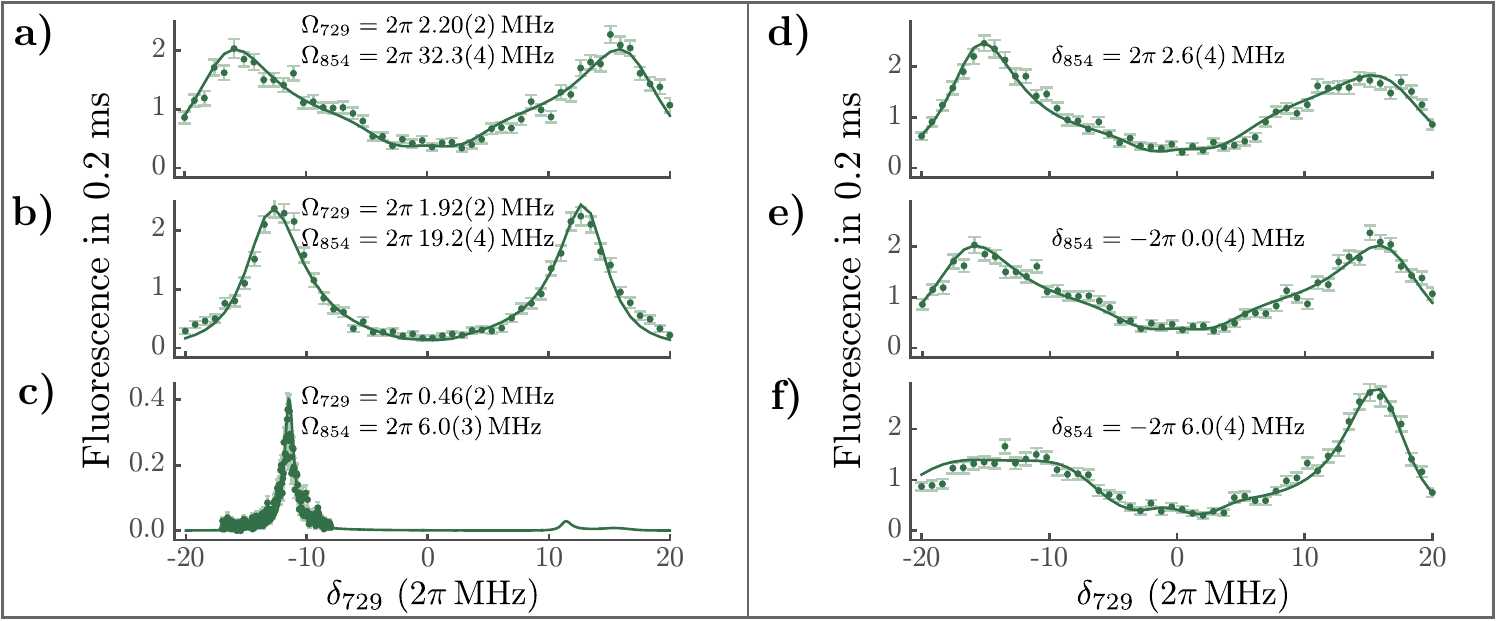}
	\caption{Left: 729\,nm/854\,nm driven fluorescence for increasing amplitudes from bottom to top and close to zero detuning of the 854\,nm laser. Right: scans with the same laser amplitudes as in panel a) but at different laser detunings $\delta_{854}$. Solid lines correspond to fits using optical Bloch equations. From these the parameters given with the plots were extracted. Each data-point was averaged over 70 experimental realizations and the standard deviation is given.}
	\label{fig:fluo393}
\end{figure}

Examples of fluorescence spectrum data and the fitted curves are shown in figure\,\ref{fig:fluo393}. The range of values obtained from the fits are within day-to-day variations which we expect in our experiment. At low laser intensities ($\Omega_{\rm 729}< 2\pi\times 1\,$MHz and $\Omega_{\rm 854} < 2 \pi \times 10$\,MHz), we were unable to fit the fluorescence data using the steady state solutions, because the internal states do not achieve the steady state in the detection time of 200\,$\mu$s. In this regime the lineshape of the measured fluorescence depends on the measurement time, and is narrower for short detection times than in the steady state. For these settings, we compared the measured fluorescence with models of integrated fluorescence predicted by the optical Bloch equations during a time of 200\,$\mu$s.

The intensities and detunings obtained from the fluorescence spectra allow us to make theoretical predictions of the cooling dynamics and the steady-state temperature. In this work we focus on the motional mode at 2.7\,MHz. To study cooling behaviour across various regimes we used the well-established control over calcium ions including pulses of 397\,nm light.
Prior to the application of 729\,nm/854\,nm laser cooling, the ion was cooled using standard Doppler cooling with UV lasers and sideband cooling on the 5\,MHz mode. The settings in the pre-cooling stage were used to adjust the temperature of the ion prior to the 729\,nm/854\,nm cooling to ensure that the ion started at a high-enough initial temperature that subsequent cooling could be observed. In cases where high intensities were used for the vis/IR cooling, the final temperature attained (and thus the initial temperature required to observe further cooling) is above the standard Doppler limit of the UV cooling. After this pre-cooling step the 729/854\,nm laser cooling was applied for a cooling time $t_c$ and then the occupancy of the motional mode at 2.7\,MHz was measured using resonant blue-sideband time scans as detailed in section\,\ref{sec:exptmethods}.

\begin{figure}
	\centering
	\includegraphics{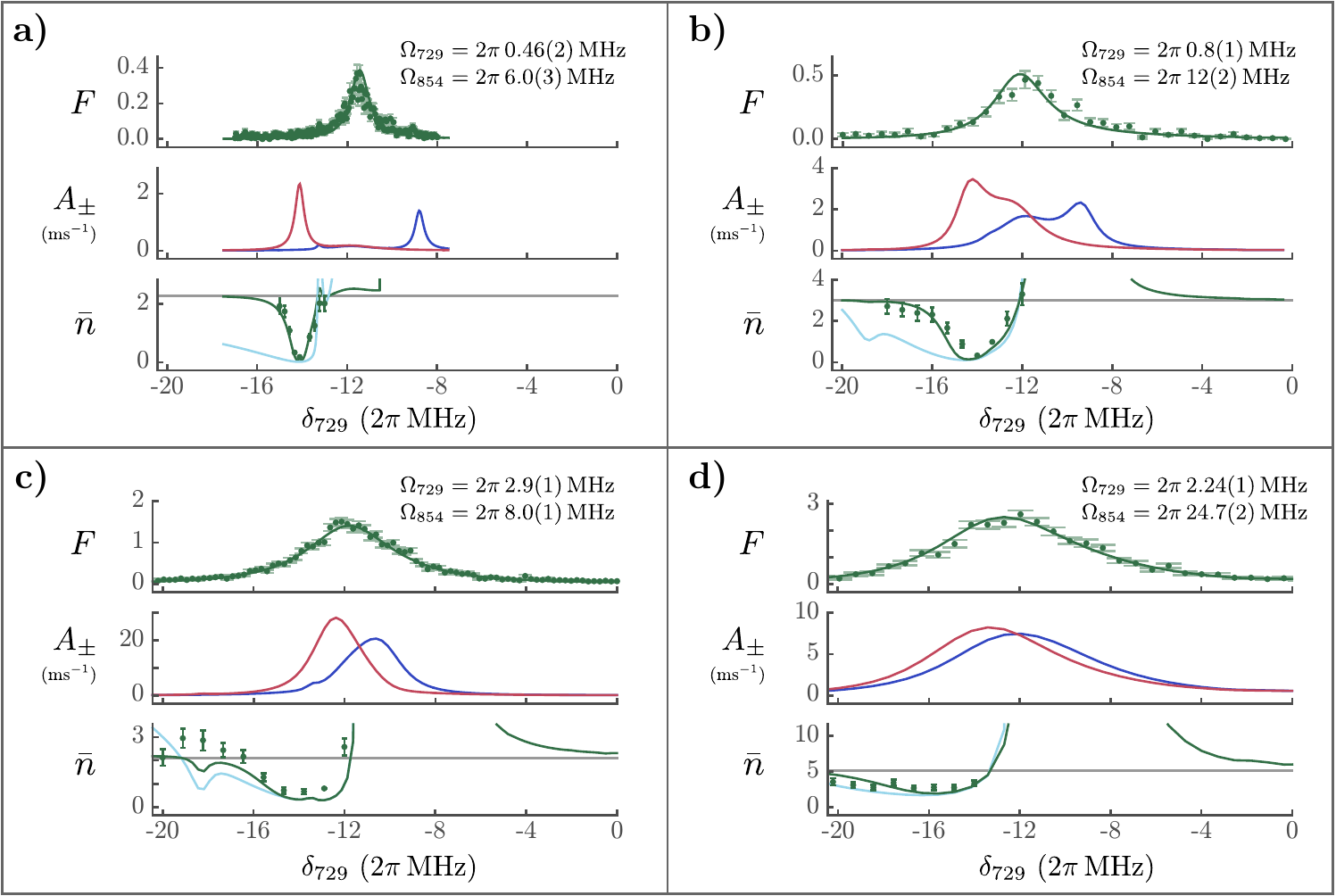}
	\caption{In the first row panels a) to d) each show measured fluorescence $F$ in a time interval of 200\,$\mu$s and a fit to it (solid line) for $\delta_{729}\leq0$. The following rows are extracted heating/cooling rate coefficients $A_\pm$ in blue/red and $\nbar$ measured against $\delta_{729}$ as well as theory curves extracted from the fit. Plotted in cyan is the predicted $\bar{n}$ for $t_c\rightarrow\infty$, in green $\bar{n}$ after $t_c$ and in grey that of $t_c=0$. For a), c) and d) $\bar{n}$ at $t_c=0$ was extracted by fitting an exponential decay to time-evolution plots shown in figure\,\ref{fig:cool3}. For b) no time-evolution plot had been taken and it was guessed from the far offresonant behaviour of $\bar{n}$.  Rabi rates extracted from the fit are indicated in the plots. For all datasets shown $\delta_{854}$ was calibrated to be close to zero with a separate experiment. Detunings extracted from the fits were (from a) to d)) $\delta_{854} = 2\pi\,4$(2) / $2(2)$ / $-3.4(4)$  / $-3.0(4)$\,MHz.
	}
	\label{fig:cool}
\end{figure}

\begin{figure}
	\centering
	\includegraphics{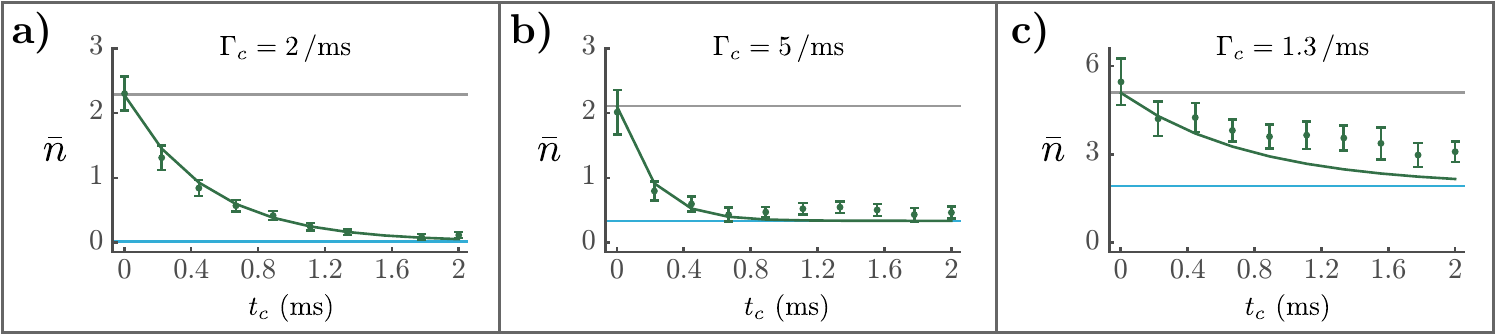}
	\caption{Time evolution of $\bar{n}$ during the cooling process. Data shown in plots a) to c) was taken under the same conditions as plots a), c) and d) in figure\,\ref{fig:cool}. The $\bar{n}$ in steady state (cyan) and after time $t_c$ (solid green line) were calculated using heating and cooling rate coefficients given in figure\,\ref{fig:cool}, giving rise to the cooling rate indicated in the plots. The $\bar{n}$ at $t_c=0$ (gray line) was extracted by fitting an exponential decay to the measured data. Data was taken at a) $\delta_{729} = -2\pi\,14.2$\,MHz, b) $\delta_{729} = -2\pi\,14.2$\,MHz, c) $\delta_{729} = -2\pi\,16.5$\,MHz corresponding to values at which we expected to reach minimal motional temperatures.
	}
	\label{fig:cool3}
\end{figure}

For fixed sets of parameters $\Omega_{\rm 729}, \Omega_{\rm 854}, \delta_{\rm 854}$ we first took a spectrum of fluorescence after ground state cooling. Then, as described above, we measured the mean vibrational quantum number of the mode as a function of $\delta_{\rm 729}$ using $t_c = 1.5$\,ms of 729/854\,nm laser cooling. Spectra and $\bar{n}$ for different fluorescence linewidths are shown in figure\,\ref{fig:cool}.
Choosing the value of $\delta_{\rm 729}$ which minimized the temperature, we also performed a further experiment in which we monitored the mean vibrational quantum number of the mode as a function of the 729/854\,nm cooling duration $t_c$ and traces are shown in figure\,\ref{fig:cool3}. Using the parameters extracted from the fluorescence trace, we can predict the heating and cooling rate coefficients $A_+$ and $A_-$ from the theory of Cirac et al.~\cite{92Cirac}; these curves are also shown in the plots. We see good agreement between the fluorescence spectrum and the measured $\bar{n}$ and cooling rates. It is clear from the predictions for the heating and cooling rate coefficients that we are able to tune continuously between the strong and weak binding limits. In the strong binding limit, $A_+$ and $A_-$ have small overlap, and are sharply peaked - this is the regime of sideband cooling. In the weak binding limit, the curves for $A_+$ and $A_-$ show strong overlap. This shows that we are able to access both the Doppler cooling and sideband cooling regimes with the same laser systems for trapping parameters which are common in quantum information experiments. In particular, the Doppler cooling regime provides a larger capture range for the cooling, which is indicated by the broader range of $\delta_{\rm 729}$ over which cooling is observed. This is helpful in the work reported in section\,\ref{sec:trapload} where cooling in this regime allows the ion to be loaded from an untrapped atomic beam, or recovered after a background gas collision.

\subsection{Trapping and re-cooling}
\label{sec:trapload}

\begin{figure}
	\centering
	\includegraphics{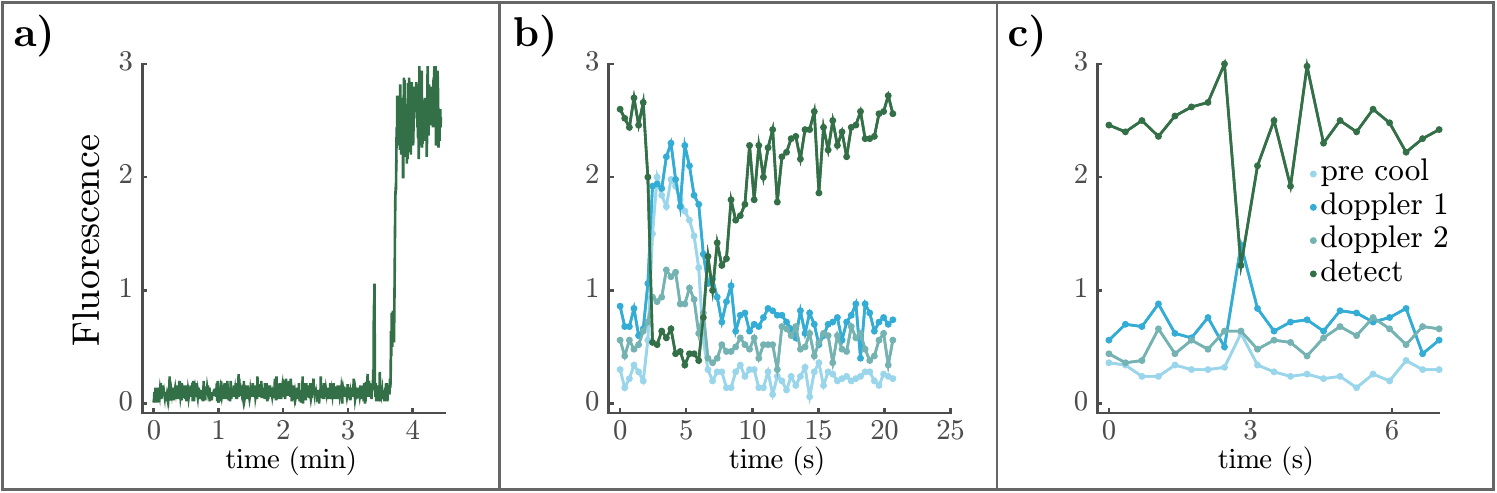}
	\caption{a) Fluorescence counts acquired in a 500\,$\mu$s detection window during ion loading. The sequence of settings given in table\,\ref{tab:sequence} were repeated throughout the scan, while 423\,nm and 375\,nm photoionization beams used to generate ions were applied continuously. b) and c) show fluorescence for re-cooling ions which have been heated up due to a background gas collision. The four traces show fluorescence detected in the different parts of the experimental sequence, using the same pulse durations, amplitudes and similar frequencies as those given in table\,\ref{tab:sequence}. The cooling intensity and detuning settings used in c) are optimized for fast re-cooling, and it can be seen that this achieves around an order of magnitude reduction in the cooling time relative to the data shown in b).}
	\label{fig:loadingrepumping}
\end{figure}

For loading ions by photo-ionization from an atomic beam, as well as for re-cooling an ion heated up by a collision with background gas atoms or molecules, a large capture range for the cooling is required. This strongly favours Doppler cooling over the sideband approach. In order to demonstrate that we can run experiments with only 729\,nm, 854\,nm and 866\,nm light, we run sequences with the settings shown in table\,\ref{tab:sequence} which do not make use of our 397\,nm laser. These were experimentally optimized to produce high scattering rates and a large cooling range. We monitor the 393\,nm fluorescence level in all parts of the experimental sequence, recording a near-background free signal by spectrally filtering background scattered light at 729\,nm, 854\,nm and 866\,nm which enters the detection apparatus with a band-pass filter centred at 390\,nm. Figure\,\ref{fig:loadingrepumping} shows data from fluorescence monitoring during both a loading event and for the recovery of an ion from two background gas collision events. Two cases of the latter are shown, before and after experimentally optimizing the cooling. While the ion is hot, fluorescence during the resonant detection phase is reduced, while it increases during the cooling periods of the sequence in which the 729\,nm laser is detuned red of resonance. This would be expected due to phase modulation of the light fields by the increased oscillation amplitude of the ion. In general, we observe that the trapping time as well as the lifetime of ions in the trap is similar using the IR/visible cooling as with the standard approach.

\begin{table}
	\begin{center}
		\begin{tabular}{l | c | c | c |c}
			& Pre cooling	& Doppler 1 	& Doppler 2 & Detection \\ \hline
			pulse time 	& 2\,ms			&2\,ms			& 1.5\,ms 	&0.5\,ms		\\
			$\delta_{729}$& -35\,MHz			& +2\,MHz				& +7\,MHz &+11\,MHz		\\
			$P_{729}$& 23\,mW			& 45\,mW				& 45\,mW &38\,mW		\\
			$P_{854}$& 25\,$\mu$W		& 25\,$\mu$W		& 25\,$\mu$W &25\,$\mu$W		\\
		\end{tabular}
	\caption{Parameters of the sequences used to load ions during the loading data shown in figure\,\ref{fig:loadingrepumping} a). The sequence runs through pre-cooling, Doppler 1, Doppler 2 and detection and then is immediately repeated.  Laser detunings and powers were optimized experimentally by minimizing re-cooling times using experimental data such as that shown in figure\,\ref{fig:loadingrepumping} b) and c). Throughout the sequence $\delta_{854}=0$.}\label{tab:sequence} 	
	\end{center}
\end{table}

\section{Simulated performance outside the range of current experiments}

The agreement between the measured fluorescence and the 12-level model within the parameter ranges of our experiment give us the confidence to use the model to predict optimal settings for high cooling and scattering rates. Below we give a summary of simulations which pursue the limits of scattering rate, cooling rate, and minimum temperatures which give an insight beyond the performed experiments.

We have seen in figure\,\ref{fig:fluo393} that detuning the 854\,nm laser does not significantly increase the maximal achievable fluorescence. The same is seen in simulations over a wider range. Keeping the detuning close to zero is favourable because then the 854\,nm light can be used to broaden the linewidth beyond what is possible with 729\,nm light alone. Therefore all simulations presented here were performed in that regime.

\subsection{Large scattering rates}

To obtain high scattering rates both 729 and 854\,nm laser Rabi frequencies should be increased. From simulations we have seen that the maximal possible scattering rate of 393\,nm photons from the ion would be $2.7\times10^7$/s, corresponding to 120\,counts/ms with our current detection system. The required Rabi frequencies on the two transitions are $\Omega_{729}\approx2\pi\times50$\,MHz and $\Omega_{854}\approx2\pi\times150$\,MHz.  This scattering rate is a factor of six above the values observed in the experiments. By comparison the maximum scattering rate in the standard 397\,nm laser cooling is $3.2 \times 10^7$/s, which gives $\approx90$\,counts/ms on our detector due to the lower detection efficiency. Although the rates are similar, the background-free nature of the detection in the case of the IR-visible fluorescence generation could be of great value, especially when high discrimination fidelity between bright and dark states is required in quantum information experiments. However here it is worth noting that the currently considered transition is not well suited for high-efficiency state detection. An alternative method is discussed in section\,\ref{sec:summary} below.

\subsection{Minimum temperature}

The lowest final temperatures within finite cooling time are reached for low driving rates. For an ion being cooled from an initial $\bar{n}=1$ for 1\,ms, simulations suggest that the lowest reachable state is $\bar{n}=0.0003$ if the only heating mechanism is off-resonant photon scattering. The optimal cooling can be reached for $\Omega_{729} = 2\pi\times0.4$\,MHz and $\Omega_{854} = 2\pi\times0.6$\,MHz. These settings are attainable in our current setup, but the lowest observed $\bar{n}$ was 0.04 quanta. The measurement itself most likely limits this number, although our trap also has a finite heating rate from electric field fluctuations at the position of the ion. This has been measured in earlier work to be 0.8\,quanta/ms at 1.8\,MHz secular oscillation frequency \cite{15Lindenfelser} - this is expected to be higher than in the current work since the secular frequency of 2.7\,MHz considered here is higher \cite{00Turchette}. The situation in which the temperature is minimized is also the setting for which the cooling rate is maximized at 120\,quanta/ms, due to the removal of a quantum of motional energy for every internal state excitation. However the capture range of sideband cooling is rather low, so these settings should be accompanied by other techniques for long ion lifetimes. 

\subsection{Large cooling capture range}

In operating a trap, one important aspect of the laser cooling is the range of Doppler shifts over which it actually provides cooling, rather than heating the ion. This is linked to the ability to catch hot ions in a trap, since hot ions tend to spend a larger fraction of time at a large Doppler shift. The efficiency of the cooling at this point is proportional to the fraction of the ion oscillation cycle duration during which it has the possibility to be cooled by the laser light. The range over which cooling can be achieved is typically closely related to the linewidths of resonances observed in the fluorescence spectrum.

In the data shown in figure\,\ref{fig:fluo393} we are able to observe fluorescence curves for which the FWHM is above 10\,MHz. By increasing the power of the 854\,nm laser beyond what was used in the experiments reported in this paper, yet within a range available in our laboratory, simulations show that it should be possible to obtain fluorescence linewidths up to 20\,MHz. Under these conditions the frequency range of the 729\,nm over which the cooling rate is over half the maximum value is 6\,MHz. Given the Zeeman substructure no simple relation between the fluorescence linewidth and the cooling linewidth or the optimal laser frequency for Doppler cooling could be found. The 2-level system result that Doppler cooling is optimally performed at a detuning equal to half the fluorescence linewidth can be taken as a good starting point for experiments though. Increasing the 854\,nm laser power too far leads to large AC~Stark shifts and results in reduced scattering and cooling rates. In the experimental data, for which we are limited by the intensity of the 729\,nm laser, cooling with a large capture range is best performed where the fluorescence levels are maximal. For $\Omega_{729} = 2\pi\times3$\,MHz this is at $\Omega_{854}=2\pi\times20$\,MHz. There the cooling rate is 5/ms (corresponding to a time-constant of 200\,$\mu$s) with a FWHM of 4\,MHz.

While not significant in the sideband cooling regime, in the Doppler cooling regime the use of co-propagating 729 and 854\,nm laser beams should lead to better cooling - this was discussed in an earlier paper by  Champenois et al. \cite{08Champenois} and observed in \cite{08Hendricks}. Using again $\Omega_{729} = 2\pi\times3$\,MHz and $\Omega_{854}=2\pi\times20$\,MHz, we simulate a cooling rate of  13\,/ms, which is a factor of $2.6$ larger than for counter-propagating beams, and a FWHM of 5\,MHz.
The availability of higher intensities in the 729\,nm laser would allow for Doppler cooling with a cooling rate of 32\,/ms (in the Lamb-Dicke regime). This rate is higher than 20\,/ms which we obtained as the maximum rate from OBE simulations of standard cooling with 397\,nm light. According to simulations of co-propagating beams an improvement on the lowest temperatures achievable in our experimental data shown in figure\,\ref{fig:cool} could only be obtained  for parameters of the data shown in figure\,\ref{fig:cool} d). There the temperature obtainable by using co-propagating beams could be 2.4 times lower.

\section{Summary and outlook}
\label{sec:summary}

The results presented above provide evidence that visible and infra-red lasers can be used to perform all of the common tasks required for quantum information experiments with trapped ions. In order to realize such experiments, it would be advantageous to also access the quadrupole transition from $S_{1/2}$ to $D_{3/2}$ at 732\,nm \cite{13Fei} along with the 729\,nm transition used in the work described above. This would allow an optical qubit based on states in the $S_{1/2}$ and $D_{5/2}$ states to be read out with high fidelity, since fluorescence detection could be performed using 732\,nm and 866\,nm light without scattering from the dark $D_{5/2}$ qubit state. We have performed simulations of the cooling cycle in this setup, and see that broadband cooling could be achieved with similar parameters to those used above. Since there is no closed three-level cooling cycle analogous to the $S_{1/2}, M_J = -1/2\rightarrow D_{5/2}, M_J = -5/2 \rightarrow P_{3/2}, M_J = -3/2$ loop used above, it could be desirable to use multi-frequency laser beams in order to obtain optimal fluorescence rates.

\ack

The authors would like to thank F. Leupold and J. Alonso for discussions in the office as well as C. Fl\"{u}hmann and U. Sol\`{e}r for helping with the 729\,nm laser system in the lab.
We acknowledge support from the Swiss National Science Foundation under Grant No. 200021\_134776, through the National Centre of Competence in Research for Quantum Science and Technology (QSIT) and
Erlangen Graduate School in Advanced Optical Technologies (SAOT).

\end{document}